\documentclass[%
reprint,
 amsmath,amssymb,
 aps,
floatfix,
]{revtex4-2}

\usepackage{graphicx}% Include figure files
\usepackage{dcolumn}% Align table columns on decimal point
\usepackage{bm}% bold math
\graphicspath{{./figures/}}

\begin{document}

\preprint{SPAM10 v2.0}

\title{Self-consistent state and measurement tomography with fewer
measurements}

% Place the author information here.  
% \affiliation command. The \affiliation command should follow the other
%information
\author{A. Stephens}
\author{J. M. Cutshall}
% \altaffiliation[Also at ]{Physics Department, XYZ University.}%Lines break
%automatically or can be forced with \\
\author{T. McPhee}
\author{M. Beck}
 \email{beckm@reed.edu}
%\author{Mark Beck}
\affiliation{Department of Physics, Reed College, Portland, Oregon, 
97202, USA}

\date{\today}

\begin{abstract}  
%\begin{singlespace}
We describe a technique for
self consistently characterizing both the 
quantum state of a single-qubit system, and the positive-operator-valued
measure (POVM) that
describes measurements on the system. The method works with only
ten measurements. We assume that a series of unitary transformations
performed on the quantum state are fully known, while making minimal
assumptions about
both the density
operator of the state and the POVM. The technique returns maximum-likely
estimates of both the density operator and the POVM. To 
experimentally demonstrate the method, we perform
reconstructions of over 300 state-measurement pairs and compare them to
their expected density operators and POVMs. We find that 95\% of the
reconstructed POVMs have fidelities of 0.98 or greater, and 92\% of the
density operators have fidelities that are 0.98 or greater.
%\end{singlespace}
\end{abstract}

\maketitle

%\doublespacing 
\section{\label{sec:intro}Introduction}
Quantum tomography is an important tool for characterizing quantum systems
and is useful for a diverse range of 
quantum information processing applications. It is useful not only for
characterizing quantum gates \cite{blume-kohout_2017,hughes_2020}, but also
for tasks
such as detecting errors in quantum key distribution
\cite{feldman_2018,moore_2020} and quantifying the randomness or privacy
of quantum-random-number generators \cite{coleman_2020,seiler_2020}.

Quantum-state tomography (QST)
estimates the density operator of an unknown quantum state by performing a
series of measurements with well calibrated detectors
\cite{vogel_1989,smithey_1993b,leonhardt_1997, altepeter_2006}.
Quantum-detector
tomography (QDT) estimates the positive-operator-valued measure
(POVM) that describes a detector, by probing it with a series of well
characterized quantum states \cite{luis_1999,fiurasek_2001,lundeen_2009}. In
quantum process tomography (QPT) the properties of
an operation that is applied to a state is characterized by operating on known
states and performing QST on the outputs
\cite{chuang_1997,poyatos_1997,dariano_2004}. 

Additionally, there exist techniques for self-consistently determining an
unknown state and an unknown measurement POVM, if one has some known state
preparations or measurements available. For example, it is possible to use
known states to calibrate detector POVMs, which are then used for QST
\cite{medford_2013}. Another option is to use a single, well-characterized
state and a limited number of high-fidelity unitary operations
\cite{keith_2018}. In data-pattern tomography one measures outcomes (data
patterns) for known states, and then matches them to outcomes for unknown
states \cite{rehacek_2010,mogilevtsev_2013,cooper_2014}. The use of
self-calibrating states is a further option \cite{mogilevtsev_2012}. 
Using somewhat different assumptions, Stark has
shown that the state and measurement operators can be determined if one has
a large set of state preparations and projective measurements (not more
general POVMs) that are globally complete \cite{Stark_2014}.

There are some techniques for self-consistently determining the
state and/or the POVM if there are no known states or POVMs. 
%; one situation that satisfies this condition is that the
%measurements are known to be projective and nondegenerate. 
% It is possible to determine both the states and the POVMs if one has a set 
% of known unitary transformations, and can prepare a single known state
% \cite{Keith_2018}. 
It is possible to perform self-characterization of quantum detectors 
without the need to know any
states \cite{zhang_2020}. Gate-set tomography
(GST) is a very general technique, where not only are the state and the POVM
determined, but so are the operators describing a series of gate operations
that are applied between the state and the measurement
\cite{blume-kohout_2013c,dehollain_2016,blume-kohout_2017,nielsen_2020}. 
Operational tomography accomplishes what GST does while using a Bayesian
framework \cite{dimatteo_2020}. 
% GST assumes that the
% Hilbert-space dimensions are known. This technique is very powerful, but a
% large number of measurements are required because there are a large number of
% parameters needed to describe the state, the measurement, and the gates.

Here we describe a technique for self-consistently estimating both the state
of a
single qubit, and the parameters of a POVM that describes a
detector, while attempting to minimize
assumptions made about the state and the POVM. Our technique is similar to that
of Ref. \cite{keith_2018} in that we assume that 
we can perform known unitary transformations between the state preparations
and the measurements.
Our technique differs from that of Ref.
\cite{keith_2018} in that we do not require any known state preparations.

The assumption that
the transformations are known will not be valid in all situations. But it
is valid if the transformations can be calibrated using a
bright, classical source and a classical detector, and this is the case 
for the polarization transformations in
our experiments (see Appendix \ref{append:pol}). Indeed, in optical
quantum information processing applications it is often the case that
unitary transformations can be calibrated classically. For example, it is
possible to implement an arbitrary linear transformation of optical modes
by using an array of 2x2 beam splitters and phase shifters
\cite{reck_1994,clements_2016}, and these transformations are now
frequently implemented using photonic integrated circuits (PICs)
\cite{carolan_2015,mennea_2018,flamini_2018}. Such circuits have been 
used to perform Boson sampling \cite{spring_2013}, teleportation
\cite{metcalf_2014}, quantum state synthesis \cite{grafe_2014}, quantum
simulation \cite{sparrow_2018}, and quantum logic operations
\cite{carolan_2015}. PICs are often characterized using classical optics
\cite{metcalf_2014,mennea_2018,guan_2019}.

% SPAMTUT is similar to GST in
% that both the state and the POVM are estimated, and in that the Hilbert space
% dimensions are assumed to be known
% \cite{[{There exist experimental techniques for determining the dimensions.
% See,
% for example,  }]Mazurek_2017}.
Both self-characterization and GST are more general than our
technique, in that they do not require known transformations
\cite{blume-kohout_2013c,zhang_2020}. However, because of
this they require more measurements. Fifty  probe states were used 
in Ref.~\cite{zhang_2020} to self-characterize the
detectors, while GST requires at least 56 measurements.
% Our technique
% requires a minimum of seven measurements, but with only seven measurements
% there are difficulties in reconstructing some states, as it may be
% necessary to divide by small numbers. 
The technique we describe here
uses 10 measurements to determine both the density operator
and the POVM describing a single-qubit system. This smaller number of 
measurements offers an advantage, even if only in time saved, to
experimentalists interested in performing these types of experiments.
% The 10 measurements are
% enough to determine an initial solution for the density operator and
% the POVM. From this solution we alternately perform maximum-likelihood
% analyses on the density operator and the POVM in order to optimize the
% final solution \cite{Keith_2018}.
% In our experiments we use individual photons encoded as polarization qubits,
% so vectors that describe the state and the POVM reside within the
% Poincar\'{e}
% sphere, and the unitary transformations act as rotations within this sphere.
% The state and the POVMs are determined to within
% a two-fold ambiguity. Essentially, this comes down to a choice of
% how one defines the basis vectors (e.g., for the polarization of individual
% photons this means defining what is horizontal and what is vertical). 
% There is
% also an undetermined continuous degree of
% freedom that leaves the magnitudes of these vectors constrained, but not
% uniquely determined. 
As is the case with most other self-consistent tomography methods,
the state and
POVM are determined to within a choice of gauge 
\cite{blume-kohout_2013c,jackson_2015,zhang_2020}. We note that operational
tomography can eliminate the need to choose a gauge, but requires some
\textit{a priori} knowledge of the system \cite{dimatteo_2020}.

\section{Theory}
\subsection{Operators and probabilities}
Suppose we have a qubit that is prepared in a state described by the density
operator $\hat \rho $, which can be expressed in terms of the Pauli matrices
${\hat \sigma _i}$ (i = 1,2,3) and the identity operator $\hat 1$ as 
\begin{equation}
	\hat \rho  = \frac{1}{2}\left( {\hat 1 + \sum\limits_{i = 1}^3 {}
	{p_i}{{\hat \sigma }_i}} \right)
    \label{eq:a}.
\end{equation}
The parameters that describe $\hat \rho $ can be arranged into a 3-component
vector $\vec p$, whose magnitude is $p$. Furthermore, we have a two-outcome
POVM described by the operators $\left\{ {\hat \Pi_1 , \hat \Pi_2 }
\right\}$,
%(They sit behind the outputs of a PBS, we assume that one fires or
%the other, but not both). 
These operators can be written in terms of 
a 3-component vector $\vec w$ (magnitude $w$) and a
bias parameter \textit{u} as \cite{jackson_2015}    
\begin{subequations}
\label{eq:b}
\begin{equation}
\hat \Pi_1  = \frac{1}{2}\left[ {\left( {1 + u} \right)\hat 1
+ \sum\limits_{i =
1}^3 {} {w_i}{{\hat \sigma }_i}} \right]
	\label{subeq:a},
\end{equation}
\begin{equation}
\hat \Pi_2  = \frac{1}{2}\left[ {\left( {1 - u} \right)\hat 1 -
\sum\limits_{i = 1}^3 {} {w_i}{{\hat \sigma }_i}} \right]
	\label{subeq:b}.
\end{equation}
\end{subequations}
These satisfy the constraint on two-outcome POVMs,  $\hat \Pi_1+ \hat \Pi_2 =
\hat 1$. Positivity is ensured by the
constraints $p \le 1$ and ${w + \left| u \right|\le 1}$.

\begin{figure}
\includegraphics[width=\columnwidth]{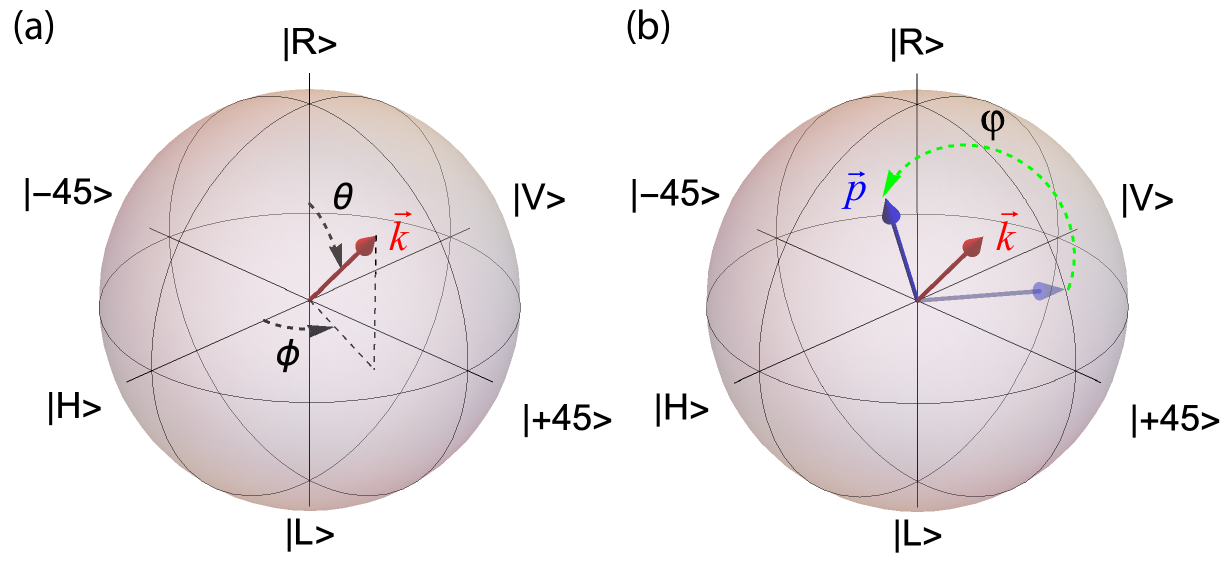}
% Here is how to import EPS art
\caption{\label{fig:1} (a) The rotation axis $\vec k$ (red) is described
in the Bloch sphere by a polar angle $\theta$, and an azimuthal
angle $\phi$. Since in our experiments we use the polarization of individual
photons as qubits, we take the 1-axis to correspond to $\left| { + 45}
\right\rangle $, the 2-axis to correspond to $\left| { R} \right\rangle $, and
the 3-axis to correspond to $\left| { H} \right\rangle $. (b) A vector that
describes the polarization state
$\vec p$ (blue) is rotated by an angle $\varphi$ about the rotation axis.}
\end{figure}

As shown in Fig. \ref{fig:1}, we define the vector ${\vec k(\theta,\phi)}$
to make
an angle of $\theta$ from the
2-axis in the 
Bloch sphere, and
its projection onto the plane perpendicular to this axis to make an angle of
$\phi$ from the 3-axis \footnote{ We define our angles in this way because in
our experiments we use qubits based on photon polarization, which is more often
depicted in the Poincar\'e sphere. In this case the 1-axis corresponds to 
+45-degree linear polarization, the 2-axis corresponds to
right-circular polarization and the 3-axis corresponds to horizontal
polarization.}.
With this convention, $\vec k$ is given by 
\begin{equation}
\vec k = \left( {\begin{array}{*{20}{c}}
{{k_1}}\\
{{k_2}}\\
{{k_3}}
\end{array}} \right) = \left( {\begin{array}{*{20}{c}}
{\sin \left( {\theta} \right)\sin \left( {\phi} \right)}\\
{\cos \left( {\theta} \right)}\\
{\sin \left( {\theta} \right)\cos \left( {\phi} \right)}
\end{array}} \right)
	\label{eq:a2}.
\end{equation}
Rotations in the Bloch sphere are given by unitary transformations 
${  {\tilde U}_j} =   {\tilde U}\left[ {{{\vec k}
(\theta_j,\phi_j)},{\varphi _j}} \right]$ 
 that are performed between the state preparation and the measurement.
 Here $j$ labels the device
settings, and we use the tilde to denote a 3x3 matrix.
This
 transformation
 rotates $\vec p$ in the Bloch  sphere by an angle 
 ${\varphi _j}$ about the axis ${\vec k(\theta_j,\phi_j)}$, and transforms it
 into $\vec {{p_j}^\prime}$: $\vec {{p_j}^\prime}=  {\tilde U}_j \vec p$.
This transformation is equivalent to $ \hat {\rho_j}' = \hat {U_j} \hat \rho$,
where $\hat {U_j}$ is the Hilbert-space operator that corresponds to the
Bloch-sphere rotation ${\tilde U}_j$.

After such a transformation, the probability that a photon
will be detected on detector 1, $ P_{1,j}$, is 
\begin{equation}
\begin{split}
{P_{1,j}} =& \;  {\rm{Tr}}\left( {\hat \Pi_1 \hat {\rho_j}' } \right)\\
%  =& \;
%  \frac{1}{4}{\rm{Tr}}\left \{ 
% {\left[ {\left( {1 + u} \right)\hat 1 + \sum\limits_{i =1}^3 {} {w_i}{{\hat
% \sigma }_i}} \right]}
% {\left( {\hat 1 + \sum\limits_{i = 1}^3 {({p_j}')}_i {{\hat \sigma }_i}}
% \right)}
%  \right \}\\
%  =& \;
%   \frac{1}{4} \Bigg [ \Bigg. (1+u){\rm{Tr}}\left( \hat 1  \right )+ 
%  \sum\limits_{i = 1}^3 w_i  {\rm{Tr}} \left( {\hat \sigma_i}   \right )+
% (1+u)\sum\limits_{i = 1}^3  {({p_j}')}_i  {\rm{Tr}}\left( {\hat \sigma_j}
% \right )\\
% &+\sum\limits_{i = 1}^3 \sum\limits_{k= 1}^3 w_i {({p_j}')}_k
% {\rm{Tr}}\left({\hat \sigma_i} {\hat \sigma_k}  \right )
%  \Bigg. \Bigg ] \\
% =&\;
%   \frac{1}{2} \left [ (1+u) + \sum\limits_{i = 1}^3 w_i {({p_j}')}_i
%  \right ]\\
=& \;
   \frac{1}{2} \left [ (1+ u) + \vec w \cdot \vec {{p_j}^\prime} \right ]\\
=& \;
   \frac{1}{2} \left [ (1+ u) + \vec w \cdot {  {\tilde U}_j}{\vec p}
 \right ] .
\end{split}	\label{eq:c}
\end{equation}
Similarly, we have
\begin{equation}
{P_{2,j}} = 
  \frac{1}{2} \left [ (1- u) - \vec w \cdot {  {\tilde U}_j}{\vec p}
 \right ] .
\label{eq:d}
\end{equation}
Assigning a value of $+1$ to a detection at 1 and $-1$ to a 
detection at 2, we can use Eqs. (\ref{eq:c}) and (\ref{eq:d})  to write 
the expectation value 
of a measurement as
\begin{equation}
 E_j= P_{1,j}-P_{2,j}=u +\vec w \cdot {  {\tilde U}_j}{\vec p}  .
\label{eq:d2}
\end{equation}

To distinguish experimentally measured and theoretically predicted
probabilities, we will use $P_{i,j}$ to represent the theoretical probability
of detection on detector $i$ for setting $j$ [Eqs. (\ref{eq:c}) and
(\ref{eq:d})], and $f_{i,j}$ to represent the corresponding experimentally
measured fraction.

\subsection{Self-consistent tomography} \label{SCT}
Our goal is to determine, in a self-consistent manner, the parameters $\vec p$,
$\vec w$ and $u$ that determine the state and the POVMs. They are determined
by applying a set of transformations ${\tilde U}_j$ and experimentally
measuring the
fractions $f_{i,j}$. An initial solution is obtained by substituting
$f_{i,j}$ for $P_{i,j}$ in Eq. (\ref{eq:d2}) and solving for $\vec p$,
$\vec w$ and $u$.

Equation (\ref{eq:d2}) is nonlinear in the components of $\vec p$ and
$\vec w$, but if we define the products of these components as 
$x_{ij}=  p_{i}   w_{j} \: (i,j=1,2,3)$ Eq. (\ref{eq:d2}) is linear in $u$
and the
nine different
$x_{ij}$'s. If we make ten measurements we can solve ten
linear equations in these ten unknowns to determine a solution for
$u$ and the $x_{ij}$'s. Details of how we do this are given in Appendix
\ref{append:eq}.

% Before we go further, we need to discuss the effects that the gauge degrees
%of
%freedom have on our solutions.
As is well described in Ref. \cite{blume-kohout_2013c}, 
self-consistent tomography techniques determine
the parameters
that describe the state and the measurements to within a choice of gauge.
State and measurement operators expressed in different gauges are 
equivalent, and every observable probability is identical. Each gauge,
therefore, corresponds to an arbitrary ``reference frame" that one must
use to express the operators mathematically
\cite{zhang_2020,blume-kohout_2013c}.
For example, by examining Eqs. (\ref{eq:c}) and (\ref{eq:d}) we see that
if we simultaneously make the
substitutions $\vec w \to -\vec w$
and $\vec p \to -\vec p$, the probabilities are unchanged. 
As such, the two solutions $(\vec w, \vec p)$
and $(-\vec w, -\vec p)$ are equivalent, but represent
different gauges and we must choose one.
% a choice must be made.
Physically, what does this
choice of gauge represent? Changing the sign of $\vec p$, for example, would
change the polarization state $\left| H \right\rangle  \to \left| V
\right\rangle $. For photons this choice of gauge effectively defines what we
mean by ``horizontal'' and ``vertical''. 
%Given it, we are able to determine constraints on $p$ and $w$. 

Next, recall that positivity places the constraints $p \le 1$ and
${w  \le 1 - \left| u \right|}$. 
Furthermore, if ${\tilde U}_1 =  {\tilde 1}$, for example, we can rewrite Eq.
(\ref{eq:d2}) as
$\vec w \cdot \vec p =wp \cos(\beta) = {E_1} - u$, where $\beta$ is the angle
between $\vec w$ and $\vec p$. 
From this we see that the measured expectation values (probabilities)
determine the product of the magnitudes of $\vec p$ and $\vec w$, 
%further subjected to the previously mentioned constraints, 
but cannot
determine their individual magnitudes because of a gauge degree of freedom.
Physically, this gauge choice trades off between the purity
of the state and the discrimination power of the detector. Because of this, 
in the reconstructions presented below we choose the gauge where
${w  = 1 - \left| u \right|}$,
and this
choice then determines $p$. 
This choice is motivated by the 
design of our experiment, where we expect that
the bias parameter is the only thing that
degrades the discriminating power of the detector \footnote{If we were using
a Bayesian analysis, we would incorporate this \textit{a priori}
information into the Bayesian prior distribution \cite{dimatteo_2020}.}. 
% this assumption is only necessary
% to compare the theoretically expected states and POVMs to the experimentally
% reconstructed ones; it is not a necessary assumption for the technique to 
% function.
% For our experimental
% apparatus we have no reason reason
% to believe that this is not the case, and the quality of our
% reconstructions from
% experimental data lend this assumption credence. 
% [ Once we have some data, we
% can hopefully say:
% As we will see below,
% making this assumption in our experiments yields very good agreement between
% the expected and measured states and POVMs.]
%Further
%\textit{a priori} information is needed to better specify $p$ and $w$. 
%Since
%we have determined all that we can about the magnitudes of $\vec p$ and $\vec
%w$, what remains is to determine their directions.

The solution to the 10 linear equations, as described in Appendix
\ref{append:eq}, determines 
$u$, which, given the discussion above, determines $w$ and $p$. What we now
need to determine is the directions of $\vec w$ and $\vec p$. Given the
measured values of the $x_{ij}$'s, if we find $p_i$, we could determine $w_j$
from $w_j=x_{ij}/p_i$. However, this is problematic if $p_i$ is 0, or nearly
so. Dividing by $w_j$ to find $p_i$ is similarly problematic. To avoid this
problem, we first find the maximum of the $x_{ij}$'s, which we refer to as 
\begin{equation}	\label{eq:h2}
    {\rm{max}} \: x_{ij}=
    x_{i_{max}j_{max}}=  p_{i_{max}}   w_{j_{max}}.
\end{equation}
With this definition, we can be confident that neither $p_{i_{max}}$ nor
$w_{j_{max}}$ are nearly 0.
We can then safely write the $w_j$'s as
%\begin{subequations}

\begin{equation} \label{eq:i}
w_j = \frac {x_{i_{max}j}}{p_{i_{max}}} .
%	\label{subeq:i:a}
% \end{equation}
% \begin{equation}
%w_2 = \frac {x_{i_{max}2}}{p_i_{max}}
%	\label{subeq:i:b}
%\end{equation}
%\begin{equation}
% w_3 = \frac {x_{i_{max}3}}{p_i_{max}}.
% 	\label{subeq:i:c}
\end{equation}
%\end{subequations}
%Thus, if we can find the $w_i$'s, we can find the $p_i$'s. 
To eliminate the $p_i$'s, we can solve these equations to write two of the
$w_j$'s in terms of the third. For example, suppose that the maximum $x_{ij}$
is $x_{23}$, so $i_{max}=2$ and $j_{max}=3$; then $p_2 = x_{23}/w_3$ and 
\begin{subequations}
\label{eq:i2}
\begin{equation}
w_1 = \frac {x_{21}}{x_{23}} w_3
	\label{subeq:i2:a}
\end{equation}
\begin{equation}
w_2 = \frac {x_{22}}{x_{23}} w_3 .
	\label{subeq:i2:b}
\end{equation}
\end{subequations}
The $w_3$ component is used to ensure that the magnitude of $\vec w$ is
consistent with our choice of gauge,
% as described above,  To find it
% we use the
% assumption that we made above: $w = 1-|u|$.
and this determines $\vec w$.
We can then solve for the components of $\vec p$ using
\begin{equation} \label{eq:i3}
p_i = \frac {x_{i{j_{max}}}}{w_{j_{max}}} .
\end{equation}

This completes the analytic solutions for $u$, $\vec p$ and $\vec w$.
% \footnote{Note that $w = 1-|u|$ is not a necessary assumption.
% Once $\vec p$ and $\vec w$ are found, we can modify their magnitudes
% subject to the constraints $x_{i_{max}j_{max}}=  p_{i_{max}}   w_{j_{max}}$,
% $w \le 1-|u|$ and $p \le 1$}.
These solutions are used as the
initial starting point for a maximum-likelihood solution.

Before moving on, note that there are seven parameters that
determine $u$ and the components of
$\vec p$ and $\vec w$. In principle one should be able to find these
parameters with only 7 measurements. Indeed, we find that it is
possible to do this for most states and POVMs. However, there are
certain special cases where a particular set of 7 measurements
might not be enough. For example, assume that $\vec p = (1,0,0)$ and
$\vec w = (0,1,0)$. In this case $x_{12}=1$, and all other $x_{ij}$'s
are 0. In order to determine $u$ and all the $x_{ij}$'s we need 10
measurements, and with only 7 measurements $x_{12}$ might not be
determined. If there are not too many 0's in the components of $\vec p$
or $\vec w$ is is possible to determine them with only 7 measurements, but
in general we cannot assume that this will be the case.

\subsection{Maximum likelihood analysis} \label{like} 
The expected probability of measuring a photon on detector $i$ for
measurement setting $j$,
$P_{i,j}$, is given in Eqs. (\ref{eq:c}) and (\ref{eq:d}). The log-likelihood
of obtaining a measured fraction of photons on this detector, $f_{i,j}$, 
is given by
\begin{equation} \label{eq:k}
%L(f_{i,j},P_{i,j})={\rm {log}}\;\mathcal{L}(f_{i,j},P_{i,j})=
L(f,P)={\rm {log}}\;\mathcal{L}(f,P)=
\sum\limits_{i,j}^{} f_{i,j}
{\rm {log}}(P_{i,j}) \; .
\end{equation}

Following Ref. \cite{keith_2018}, we maximize the likelihood
function by alternating between QST and QDT. When performing QST we hold
the POVM fixed, and the density operator is fixed while performing QDT.

For QST we use the $R \rho R$ method \cite{Rehacek_2007,Glancy_2012}. In
this method the density
operator at iteration $k+1$ is written in terms of the density operator at
iteration $k$ as
\begin{equation} \label{eq:l}
    {\hat \rho}^{(k+1)}= \hat R^{(k)} {\hat \rho}^{(k)}
    \hat R^{(k)}  \; ,
\end{equation}
where 
\begin{equation} \label{eq:m}
    \hat R^{(k)} = \sum\limits_{i,j}^{} \frac{f_{i,j}}{P^{(k)}_{i,j}}
    \hat \Pi_{i,j} \; 
\end{equation}
and $\hat \Pi_{i,j}=\hat \Pi_i \hat U_j$. At each
iteration the likelihood is guaranteed to increase, and the density operator is
guaranteed to remain positive. After each iteration we renormalize the density
operator.

For QDT we use the method of Lagrange multipliers described in Refs.
\cite{fiurasek_2001,chen_2019}. The POVM in iteration $k+1$ is given by
\begin{equation} \label{eq:n}
    {\hat \Pi}^{(k+1)}_{i}= \hat R'^{(k)}_i {\hat \Pi}^{(k)}_i
    \hat R'^{(k)}_i ,
\end{equation}
where 
\begin{equation} \label{eq:o}
    \hat R'^{(k)}_i =\sum\limits_{j} \frac{f_{i,j}}{P^{(k)}_{i,j}}
    \left (\sum\limits_{l}\sum\limits_{m,n}^{}
    \frac{f_{l,m}}{P^{(k)}_{l,m}} \frac{f_{l,n}}{P^{(k)}_{l,n}} {\hat \rho}_m
    {\hat \Pi}^{(k)}_l
    {\hat \rho}_n \right )^{1/2} \hat \rho_j \; .
\end{equation}
Here $\hat \rho_j=\hat U_j \hat \rho$. Again, the likelihood is guaranteed
to increase after each iteration, and the POVM will always be positive.

We use separate termination conditions for QST and QDT, stopping when both
conditions are reached. To set the stopping point for QST, we follow the
method of Refs. \cite{keith_2018,Glancy_2012}. At each iteration we calculate
$S_{\rho}$, which is an upper bound on the difference between the current
likelihood and the unknown maximum likelihood
\begin{equation}
\begin{split}
L\left({\hat \rho}_{ML}\right) - L\left({\hat \rho}_k\right) \leq&
\text{max}\left[\text{eigenvalues}\left(\hat{R}^{(k)}\right)\right] - N \\
=& S_{\rho}.
\end{split}
\end{equation}
Here ${\hat \rho}_{ML}$ is the density operator at maximum likelihood and $N
$ is the number of measurements, in our case 10.

To set the stopping point for QDT we use the Frobenius norm of consecutive
iterations of the POVM. At each iteration, we calculate
\begin{equation}
\left|\left|{\hat \Pi}^{(k)}_{1} - {\hat \Pi}^{(k+1)}_{1}\right|\right|
= S_{\Pi}.
\end{equation}
We stop once both $S_{\rho}$ and $S_{\Pi}$ are less than $10^{-5}$. 

It would be possible to use other, possibly more efficient, techniques for
maximizing the likelihood. But we have found that the technique we use works
well, and is more than efficient enough (the experimental data is acquired in
minutes, while the data analysis takes only seconds).

\subsection{Figures of merit}
% Once we have found $\vec p$, $\vec w$ and $u$, we can use Eqs. (\ref{eq:a})
% and
% (\ref{eq:b}) to find $\hat{\rho}$ and the $\hat{\Pi}$'s. 
To compare the
theoretically expected POVM (or state) to that reconstructed by our
technique we
use the fidelity $F$, which for two POVMs is given by
\cite{jozsa_1994,zhang_2012}
\begin{equation}
F= \frac {{\left[ {{\rm{Tr}}\left( {\sqrt {\sqrt {\hat \Pi_1} \hat \Pi_2\sqrt
{\hat \Pi_1 } }} \right)} \right]^2}}
{{\rm{Tr}}({\hat \Pi_1}){\rm{Tr}}({\hat \Pi_2})}. \label{eq:p}
\end{equation}
The fidelity takes on values $0 \le F \le 1$ , with $F=1$  corresponding to
${\hat\Pi_1}={\hat\Pi_2}$ and $F=0$ corresponding to orthogonal 
operators. To
compare density operators we simply replace $\hat{\Pi}$ by $\hat{\rho}$. 

The fidelity is a convenient figure of merit in that it is a measure of how
well the measured state agrees with the expected state. However, one needs
to trust one's knowledge of the expected state in order to have confidence
in the fidelity. As such, it is also convenient to have a measure that does not
depend on explicit knowledge of the expected state. Here we use the
total variation distance (TVD) to compare the experimentally measured
fractions $f_{i,j}$ to the probabilities returned by the fit to the model,
$P_{i,j}$ \cite{rudinger_2019}. The TVD is a measure of how close the
measured and modeled probabilities are, and it is given by
\begin{equation} \label{eq:p2}
{\rm{TVD}} = \frac {1}{2} \sum \limits_{i,j}| P_{i,j}-f_{i,j}| .
\end{equation}

The TVD also has another useful property. Since it depends only on measured
and modeled probabilities, and these are independent of the choice of
gauge, the TVD is gauge invariant. Any choice of gauge that we
make will not affect our determination of the TVD. However, there is no
well-motivated measure of fidelity that is gauge invariant
\cite{blume-kohout_2013c}. As such, in our experiments we choose between
the $(\vec w, \vec p)$
and $(-\vec w, -\vec p)$ gauges by using the one that maximizes the fidelity
with the expected state \footnote{For this particular gauge degree of
freedom, where we have only two
possible gauges to chose from, and fixing the gauge is only necessary to
calculate the fidelity, we find that maximizing the fidelity
works sufficiently well.}.  
\subsection{Larger numbers of qubits}
The potential exists for scaling this technique to larger numbers of qubits. An
efficient means for doing this is as follows. Imagine a 2-qubit system with two
detectors. 
We can simply use the technique above to perform detector tomography
on each of the two detectors, even without knowing the state.
This requires 10 measurements to be performed with
each detector. We can then use the reconstructed POVMs to perform QST on the
source. For a
2-qubit system, QST can be accomplished with as few as 9 measurements 
\cite{altepeter_2006}. As
such, we can self-consistently determine both the unknown state, and both
POVMs for
a 2-qubit system, with $\sim\!30$
measurements. Due to the gauge degrees of freedom, there
are 4 possible state-POVM pairs that describe the system (the continuous
gauge degrees of freedom are also still present). 

Note, however, that if the source is in a Bell state, for example, the marginal
density operator for each individual qubit is a perfectly random mixed
state. In
this case all of the $x_{ij}$'s are 0, and the technique described above will
not work. As such we need to use conditional measurements on one qubit to
project the other qubit into a state that is not perfectly random. In this way
we can perform detector tomography on each detector. Note that conditional
measurements are $only$ needed if the marginal distributions for the one or
both of the qubits
are perfectly random.

No knowledge of the underlying state is needed here. 
For example, consider the Bell state
\begin{equation}
{\vert{\phi^+}\rangle} = \frac {1}{2} \left( {\vert{H,H}\rangle} + {\vert{V,V}\rangle} \right) .
\end{equation}
A measurement of polarization on one of the photons will project it onto some
elliptical polarization state ${\vert{e_1}\rangle=a\vert{H}\rangle + b
e^{i\phi}\vert{V}\rangle}$. This projects the other photon into the
polarization state ${\vert{e_2}\rangle=a\vert{H}\rangle + b
e^{-i\phi}\vert{V}\rangle}$ via quantum steering. If measurements of the
second photon are performed conditionally with the first, the second
photon will not
be in a random mixed state, and it will be possible to reconstruct both its
state and the POVM of the detector that performs measurements on it. If the
original beam is in a different Bell state then the second photon will be
projected into a different state, but it will always be the case that it will
not be in a perfectly random mixed state. 

For other more general two-qubit states whose marginals are perfectly
random, any conditional measurement
that projects the conditionally prepared state into a state that is not
perfectly random will suffice. As long as there is some correlation between
the qubits, nearly any measurement should accomplish this.

\section{Experiment}
\subsection{Apparatus}

We use 5~mW of power from a 405~nm, single-frequency laser diode to pump a
25~mm long, type-II,
periodically-polled, potassium titanyl phosphate (PPKTP) crystal. This
produces spontaneous-parametric-down conversion
at 810~nm, and we separate signal and idler beams with a polarizing beam
splitter (PBS). The idler beam is focused into a single-mode
optical fiber, filtered by
a 10~nm bandwidth filter centered at 810~nm, and detected by a
single-photon-counting module (SPCM). Detection of an idler photon heralds the
production of a coincident, single photon in the signal beam. For coincidence
counting we use a coincidence window of 3.2 ns on a commercial
time-to-digital converter, and we subtract the expected 
accidental coincidences.

\begin{figure}
\includegraphics{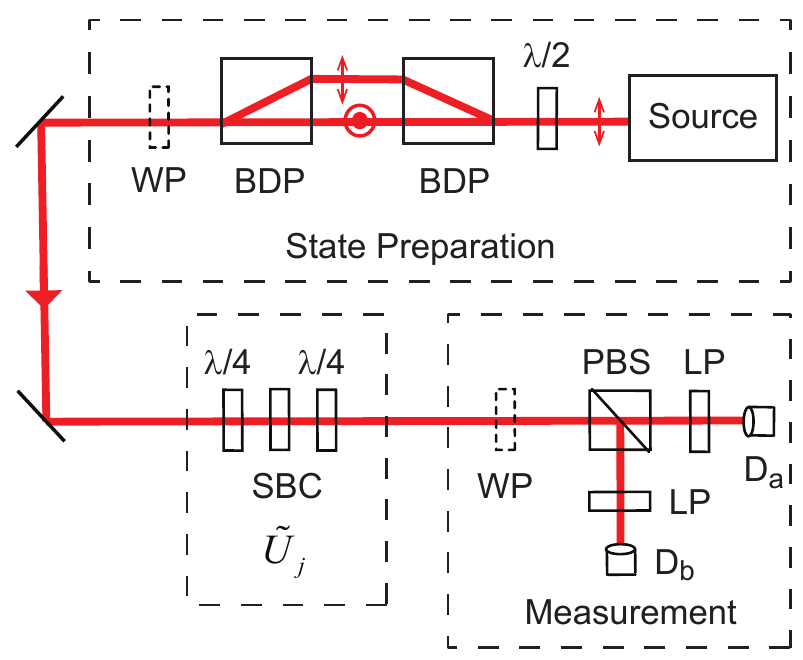}% Here is how to import EPS art
\caption{\label{fig:app} The experimental apparatus. The source consists of
heralded single photons emerging from a single-mode optical fiber.
BDP denotes a beam-displacing polarizer,
$\lambda / 2$ denotes a half-wave plate, $\lambda / 4$ denotes a quarter-wave
plate, WP denotes optional half- and/or quarter-wave plates, SBC denotes a
Soleil-Babinet compensator, LP denotes a linear polarizer and PBS denotes
a polarizing beam splitter (a
Rochon prism). Light emerging from the LPs is wavelength filtered and coupled 
into multi-mode
optical fibers (not shown), which are then coupled to single-photon-counting
modules (labeled ${D_a}$ and ${D_b}$). }
\end{figure}

The signal beam is 
focused into a
single-mode, polarization-preserving optical fiber, and emerges as
the ``Source” in Fig.~\ref{fig:app}. Before being detected with SPCMs,
signal photons are
also filtered with 10~nm bandwidth, 810~nm filters. The heralded signal
photons produced by our source have a measured degree of second-order
coherence $g^{(2)}(0)=0.024 \pm 0.002$, so the signal beam is well
described by a
single-photon state. Furthermore, by blocking the signal beam we find
that the ratio of heralded 
background detections, including dark counts, to heralded signal-photon
detections is 0.0004. As such, we conclude that background detections
have a minimal effect on our measurements.

Linearly-polarized photons from the source pass through a half-wave plate that
rotates their polarization. These photons then pass through a beam-displacing
polarizer (BDP) that spatially displaces the horizontal component of the
polarization from the vertical component; the fraction of the horizontal and
vertical components is adjusted by the rotation angle of the half-wave plate.
A second BPD spatially recombines the beams, but the horizontal component is
delayed by a time longer than the coherence time of the individual photons.
This creates an adjustable mixture of horizontal and vertical polarizations.
Half- and/or quarter-wave plates placed after the BDPs allow us to rotate the
polarization, and create any state of single-photon polarization.
Likewise, half- and/or quarter-wave plates in front of a PBS allow us to
perform measurements of any projection of the
polarization. 
%In Eq.~(\ref{eq:b}) we assume that our two detectors are
% described by POVMs that have the same parameters. To ensure that the detector
% efficiencies are the same, we insert linear polarizers after the PBS. These
% allow us to adjust the amount of light hitting the two detectors, and hence
% balance their efficiencies. 

In our data analysis we are able to model the POVMs that describe our
detectors in two different ways, and we will present the results separately.
In the first model we treat the entire subsystem labeled ``Measurement"
in Fig.~\ref{fig:app} as a two-outcome POVM. We post select on
coincident detections between an idler photon, and a signal photon at
either $D_a$ or $D_b$. In this model detection at $D_a$ corresponds to
${\hat \Pi}_1$ and detection at $D_b$ corresponds to ${\hat \Pi}_2$. We
exclude
events with heralded detections at both $D_a$ and $D_b$, which are small
in number because of our low measured value of $g^{(2)}(0)$.
As described above,
background events are a small percentage of coincidence detections, so
the vast majority of our post-selected events represent true
signal-photon detections, and include only a very small number of events where
no photons were present at $D_a$ or $D_b$.

In order to treat the two detectors as corresponding to different operators
in a two-outcome POVM, it is necessary for their corresponding values of
$u$ and $\vec w$ to be the same. The use of a high-quality Rochon PBS (extinction ratio of $>\!10^4$ for both polarizations) ensures
that the two detectors monitor orthogonal polarizations, thus ensuring
that $\vec w$ is the same. To ensure that $u$ is the same we need
the detection
efficiencies to
be the same. We do this by inserting
linear polarizers after the PBS. These
allow us to adjust the amount of light hitting the two detectors, and hence
balance their their measured count rates to within $\sim3\%$.
% In this model we do not expect any reduction
% of the discrimination power of the
% detector, which would reduce $w$, other than that which might result from
% some imbalance at the beam splitter described by $u$. 
% Therefore, we
% assume $w=\nobreak 1- \left
% |u \right |$ during our reconstructions. This provides enough information to
% eliminate the ambiguity due to the continuous gauge degree of freedom, and
% hence uniquely determine $p$ from the data. 
The advantage of this
post-selected detector model is that inefficiencies in detection do not effect
the POVMs.
%This design also ensures that our choice of gauge, $w=1-|u|$, is
%reasonable. 

In the second model, each detector is assigned its
own two-outcome POVM. For example, consider $D_a$: ${\hat \Pi}_1$ corresponds
to a coincident detection between the idler detector and $D_a$, while
${\hat \Pi}_2$ corresponds to a heralding detection at the idler detector,
but no coincident detection at $D_a$. The heralding serves as a ``clock" that
tells us when to interrogate the detector and observe the outcome.
This model has the advantage that we
need not assume that the values of $u$ and $\vec w$ for $D_a$ and $D_b$ are
the same.
The disadvantage is that for our relatively low heralding efficiency,
the imbalance $u$ between no detections and detections is large, and
dominates the parameters of the POVM.

We use a Soleil-Babinet compensator (SBC) placed between two quarter-wave
plates to implement the transformations
$ {\tilde U}\left[ {{{\vec k}
(\theta_j,\phi_j)},{\varphi _j}} \right]$. 
The rotation angles of
the wave plates and the phase shift of the SBC that make up 
these transformations are all under computer control. 
We use the theoretically expected
$\tilde U_j$'s, given the settings of our device, to perform our
tomographic reconstructions. Our calibrations,
details of which are found in Appendix \ref{append:pol}, 
show that all of our $\tilde U_j$'s have a mean process fidelity of
at least 0.994 with the actual experimentally implemented transformation.

To generate the
10 measurements necessary to determine the state and the detector POVMs 
the computer
steps through the transformations and records the singles and coincidence
counts. For each measurement setting we acquire approximately 20,000
coincident detections. From the raw counts we calculate the probabilities
and expectation values necessary to perform the reconstructions. 

% Since we look for coincidences between a signal and idler photon, counts for
% the B and Bp detectors are denoted $N_{AB}$ or $N_{ABp}$ respectively. Since
% we consider both detectors simultaneously, the expectation value is
% calculated by comparing the counts on each,
% \begin{equation}
%     E= \frac{N_{AB}}{N_{AB}+N_{ABp}}-\frac{N_{ABp}}{N_{AB}+N_{ABp}}.
% \end{equation}

\subsection{Results}

\begin{figure}
    \centering
    \includegraphics[width=\columnwidth]{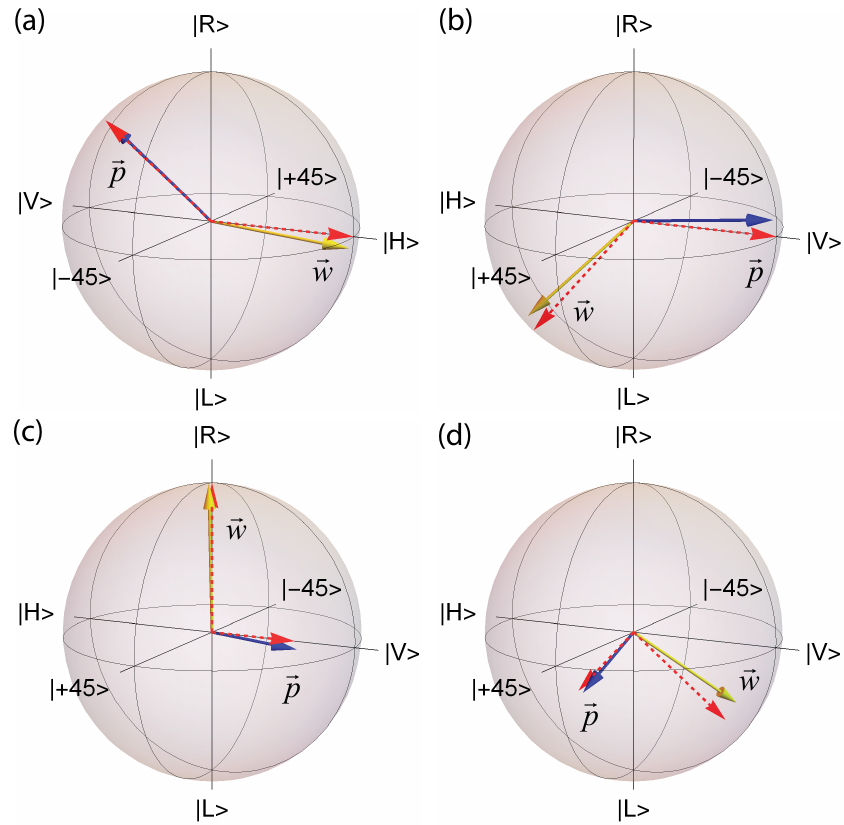}
    \caption{Experimental determinations of the vectors that
describe the state $\vec p$ (blue) and the POVM $\vec w$ (yellow); (a)-(d)
show four different experiments. The corresponding theoretically expected
vectors are shown as dashed red arrows. The reconstructions here use the
first detector model.}
    \label{fig:bloch}
\end{figure}

\begin{table*}
\caption{\label{tab:table1}The data corresponding to the plots in Fig.
\ref{fig:bloch}.}
\begin{ruledtabular}
\begin{tabular}{ccccccc}
% &\multicolumn{2}{c}{$D_{4h}^1$}&\multicolumn{2}{c}{$D_{4h}^5$}\\
Fig.&$\vec p$ Expected&$\vec p$ Measured&$\vec w$ Expected&$\vec w$
Measured &$u$ Expected&$u$ Measured\\ \hline
\ref{fig:bloch}(a)&$(-1/2,1/\sqrt{2},-1/2)$&$(-0.486,0.697,-0.488)$
&$(0,0,1)$&$(0.017,-0.084,0.970)$&0&0.026 \\
\ref{fig:bloch}(b)&$(0,0,-1)$&$(0.057,0.117,-0.991)$&$(1/2,-1/\sqrt{2},1/2)$&
$(0.487,-0.620,0.553)$&0&0.037\\
\ref{fig:bloch}(c)&$(0,0,-0.6)$&$(-0.076,-0.074,-587)$&$(0,1,0)$&
$(0.045,0.998,0.0004)$&0&0.0006\\
%(0.275,-.389,0.275)= (1/2,-1/sqrt(2),1/2)*0.55
\ref{fig:bloch}(d)&$(0.275,-.389,0.275)$&$(0.356,-0.364,0.201)$&
$(-1/2,-1/\sqrt{2},-1/2)$&$(-0.571,-0.588,-0.556)$&0&0.009\\
\end{tabular}
\end{ruledtabular}
\end{table*}

\begin{table}
\caption{\label{tab:table2}The fidelities and total variation distances
corresponding to the data in Fig. \ref{fig:bloch} and Table \ref{tab:table1}.}
\begin{ruledtabular}
\begin{tabular}{ccccc}
% &\multicolumn{2}{c}{$D_{4h}^1$}&\multicolumn{2}{c}{$D_{4h}^5$}\\
Fig.&Fidelity $\hat{\rho}$&Fidelity ${\hat\Pi}_1$&Fidelity ${\hat\Pi}_2$&
TVD\\ \hline
\ref{fig:bloch}(a)&0.990&0.998&0.998&0.065 \\
\ref{fig:bloch}(b)&0.996&0.998&0.998&0.073 \\
\ref{fig:bloch}(c)&0.997&0.988&0.999&0.124 \\
\ref{fig:bloch}(d)&0.997&0.994&0.994&0.049 \\
\end{tabular}
\end{ruledtabular}
\end{table}

We have performed measurements for different states and detector POVMs.
Essentially, we place $\vec p$ and $\vec w$ in 
different places in the Bloch sphere. We vary the directions of both of
these vectors, and we also vary the magnitude of $\vec p$ by controlling the
purity of the state. 
%Recall from Sec. \ref{SCT} that we assume $w=1-|u|$.
%The detectors were balanced so that $u\approx0$. 
We perform five trials for each set of parameters that determine the state
and the POVMs. We performed a
total of 310 trials. 

\subsubsection{First detector model}
Four example reconstructions are
shown in Fig.~\ref{fig:bloch}. In this figure we are using the first
detector model, in which detection at $D_a$ corresponds to
${\hat \Pi}_1$ and detection at $D_b$ corresponds to ${\hat \Pi}_2$.
% Fig.~\ref{fig:bloch}(a) shows a trial where
% $\vec w$ points along the horizontal axis and $\vec p$ (a nearly pure state)
% is off the equator. The expected magnitudes are p=1 and w=1, while the
% measured magnitudes are $p=0.980$, $w=0.974$, and $u=0.026$. 
The theoretically expected, and experimentally determined parameters that
describe the state and the POVMs corresponding to those displayed in
Fig.~\ref{fig:bloch} are given in Table \ref{tab:table1}. The
theoretically expected parameters are calculated from the known
wave-plate settings that determine the state and the measurement.

From the reconstructed state and POVM parameters we can use Eqs.
(\ref{eq:c}) and
(\ref{eq:d}) to calculate the detection probabilities $P_{i,j}$ associated
with the model. These and the measured fractions $f_{i,j}$ determine the
TVD of the reconstruction using Eq. (\ref{eq:p2}). 
% Using 
% Eqs.~\ref{eq:a} and~\ref{eq:b}, we can calculate the density matrix
% $\hat \rho$
% and POVMs ${\hat \Pi}_i$ that correspond to the theoretical predictions and
% the experimental reconstructions. 
We can use Eq.~(\ref{eq:p}) to calculate
the corresponding fidelities. The TVDs and fidelities corresponding to
the entries in Table \ref{tab:table1} are given in Table \ref{tab:table2}.

\begin{figure}
    \centering
    \includegraphics{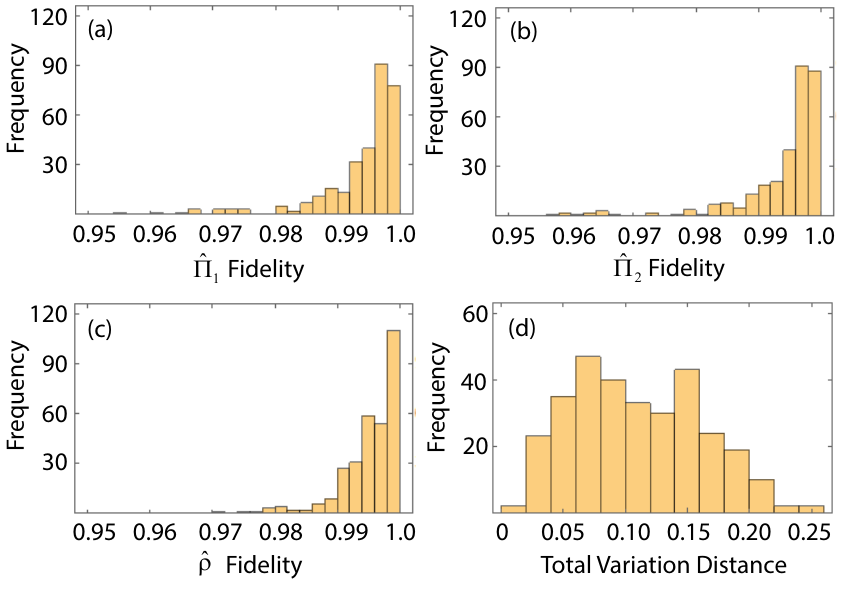}
    \caption{Histograms (a) and (b) show the fidelities of
    the theoretically expected ${\hat \Pi}_1$ and ${\hat \Pi}_2$ with the
    experimentally determined operators, (c) shows the fidelity of the
    theoretical and experimental $\hat \rho$, and (d)  the total variation
    distance between the measured and the model probability distributions.
    This data was analyzed using the first detector model.}
    \label{fig:hists}
\end{figure}

The fidelities of $\hat \rho$ and the ${\hat \Pi}_i$'s, and the total variation
distances of all the trials are shown as histograms in Fig.~\ref{fig:hists}.
%This data for this figure comes from the first detector model.
For the POVMs the fidelities are above
0.99 for 82\% of trials, and above 0.98 for 95\% of the trials.
Fidelities for the density matrices are above 0.99 in 91\% of trials,
and above  0.98 in 98\% of trials. The mean TVD for all trials was found to be
$0.11 \pm 0.05$. There are 20 terms in the sum of Eq. (\ref{eq:p2}) for
the TVD (2 outcomes, 10 measurements), so on average the modeled
probabilities and the measured fractions differ by approximately 0.01.
% It
% should be noted that above we said that we are assuming that $w=1-|u|$
% in order to uniquely specify the magnitudes of $\vec w$ and $\vec p$.
% This assumption is necessary to compare the theoretically expected
% and the experimentally reconstructed parameters and calculate, for
% example, fidelities. However, this assumption does $not$ effect the
% TVD. Our method uniquely determines the product of the magnitudes $pw$,
% where the magnitudes are further constrained by $w \le 1-|u|$ and
% $p \le 1$. Any solution that satisfies all of these constraints will have
% the same TVD, as all solutions predict the same values for the
% probabilities.

\begin{figure}
    \centering
    \includegraphics{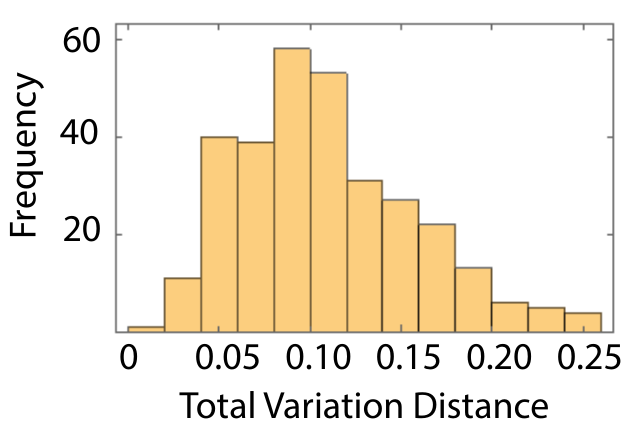}
    \caption{Histogram of TVDs obtained from numerically simulated data, 
    using parameters chosen to align with those of the experimental TVDs
    displayed in Fig. \ref{fig:hists}(d). }
    \label{fig:SimHists}
\end{figure}

We have estimated our experimental ability to accurately generate 
the Bloch sphere transformations
${\tilde U}\left[ {{{\vec k}
(\theta_j,\phi_j)},{\varphi _j}} \right]$
by comparing our experimentally measured TVD data to numerically simulated
data.
We performed simulations while varying the amount of error 
in the transformation angles, and compared the resulting TVDs to the
distribution in
Fig.~\ref{fig:hists}(d). The statistics of the errors were assumed to be
the same for each of the three angles, having a Gaussian distribution with 0
mean and an adjustable standard deviation. Statistical errors due to
finite numbers of counts were also simulated. 
After performing simulations with 
differing amounts of error, we find that the
distribution of TVDs in the simulations were most similar to those of
Fig.~\ref{fig:hists}(d) for a standard deviation of 
0.05 radians, and the simulated TVDs in this case are shown in Fig.
\ref{fig:SimHists}. For this amount of error, the simulated TVD's had a mean
of $0.11 \pm 0.04$. 
%The distribution of experimental TVDs is slightly
% ``flatter", indicating that the experimental errors may not be perfectly
% described by a Gaussian distribution, but otherwise the agreement is quite
% good. 
The agreement between the simulations and the experiment lead us
to believe that the accuracy of our technique is currently limited
by errors on the order of 0.05 radians in controlling the angles in the
transformations
${\tilde U}\left[ {{{\vec k}
(\theta_j,\phi_j)},{\varphi _j}} \right]$ that we desire
\footnote{The calibration of the unitary transformations
presented in
Appendix \ref{append:pol} yields fidelities consistent with those of
Fig.~\ref{fig:hists}(c). This lends further credence to the belief that
small errors in in the transformations limit the accuracy of our
reconstructions.}. It might be possible to improve our results by
applying a neural
network or a Bayesian analysis to help
us compensate for these errors \cite{dimatteo_2020,palmieri_2020}.

\subsubsection{Second detector model}
\begin{table*}
\caption{\label{tab:table3}The data for the second detector model, as
applied to detector $\rm{D_b}$, and 
corresponding to the same measurements as in TABLE \ref{tab:table1}.}
\begin{ruledtabular}
\begin{tabular}{cccccc}
% &\multicolumn{2}{c}{$D_{4h}^1$}&\multicolumn{2}{c}{$D_{4h}^5$}\\
$\vec p$ Expected&$\vec p$ Measured&$\vec w$ Expected&$\vec w$ Measured &
$u$ Expected&$u$ Measured\\ \hline
$(-1/2,1/\sqrt{2},-1/2)$&$(-0.482,0.694,-0.482)$&$(0,0,-0.046)$&
$(-.001,0.004,-0.046)$&-0.960&-0.954 \\
$(0,0,-1)$&$(0.057,0.089,-0.972)$&$(-0.025,0.035,-0.025)$&
$-0.027,0.033-0.028)$&-0.960&-0.949\\
$(0,0,-0.6)$&$(-0.060,-0.068,-0.558)$&$(0,-0.035,0)$&
$(-0.002,-0.035,-0.001)$&-0.960&-0.964\\
$(0.275,-.389,0.275)$&$(0.329,-0.347,0.194)$&$(0.017,0.023,0.017)$&
$(0.020,0.019,0.019)$&-0.960&-0.967\\
\end{tabular}
\end{ruledtabular}
\end{table*}

\begin{table}
\caption{\label{tab:table4}The fidelities and total variation distances
corresponding to the data in TABLE
\ref{tab:table3}.}
\begin{ruledtabular}
\begin{tabular}{ccccc}
% &\multicolumn{2}{c}{$D_{4h}^1$}&\multicolumn{2}{c}{$D_{4h}^5$}\\
Fidelity $\hat{\rho}$&Fidelity ${\hat\Pi}_1$&Fidelity
${\hat\Pi}_2$&TVD\\ \hline
0.986&0.998&0.999996&0.0042 \\
0.986&0.999&0.999991&0.0066 \\
0.997&0.999&0.999998&0.0030 \\
0.997&0.993&0.999675&0.0029 \\
\end{tabular}
\end{ruledtabular}
\end{table}

In the second detector model each detector is treated separately, and
the two-outcome POVMs correspond to heralded detections
${\hat \Pi}_1$ or no detections ${\hat \Pi}_2$
for each detector. We use the same experimental data as was used in
the first detector model, we just analyze it differently.  

Expected and reconstructed state and measurement parameters for
detector $D_b$ are shown in Table \ref{tab:table3}
(all results for detector $D_a$ are very similar to those of $D_b$), while
Table \ref{tab:table4} shows the measured fidelities and TVDs corresponding
to this data. Each row of Table \ref{tab:table3}
corresponds to an analysis of the same data used to construct
the corresponding row of Table \ref{tab:table1}. The primary difference
between the two models is the imbalance parameter $u$. The first model
is insensitive to the heralding efficiency, while the second model is
sensitive to it. The heralding efficiency is equal to
$1-|u|$, and we see that the data in Table \ref{tab:table3} reflect an
overall heralding efficiency of 3-5\%, which is consistent with
independent measurements of this parameter. We measure this efficiency 
by dividing the total number of signal-idler coincidence detections by
the total number of idler (heralding beam) detections \footnote{We note that
there are more sophisticated techniques for calculating the efficiency that
take into account imperfections in the measurements
\cite{ma_2007,takesue_2010}. Since our measurements
for $g^{(2)}(0)$ and the background count rate are small, these corrections would
also be small.}. We find that the efficiency
fluctuates slightly
from day to day as the fiber-coupling of the source changes.
The low efficiency is
primarily due to the fact that we have not optimized the coupling
of our source into the single-mode fibers. 

In Table \ref{tab:table3} the theoretically expected parameters that
describe $\vec p$ and $\vec w$ are calculated from the known
wave-plate settings that determine the state and the measurement.
%, and 
%Table \ref{tab:table4} shows the measured fidelities and TVDs corresponding
%to this data.
% However, we cannot theoretically predict $u$ with any degree of precision,
% and it must be measured.
% The most accurate measurement of $u$ for any data set
% is given by the solution to Eq.
% (\ref{subeq:g:a}), which is the average of 4 measured expectation
% values from the data. This is also the initial solution for $u$
% that we seed the maximum-likelihood analysis with. As such, the
% theoretically expected value of $u$ in Table \ref{tab:table3} is
% the initial solution before the maximum-likelihood analysis, while
% the experimentally measured value is that returned by this analysis.
% In this sense these two values are not independent of each other.
% Having said that, the fidelities in Table \ref{tab:table4} are not especially
% sensitive to small changes in the expected values of $u$. If this value is 
% varied between -0.93 and -0.98, the fidelities for ${\hat \Pi}_1$ in Table
% \ref{tab:table4} are unchanged, while those of ${\hat \Pi}_2$ remain above
% 0.999. The values of $u$ as calculated from our
% independent measurements of the heralding efficiency, as described above, lie
% within this range. Thus,
% recalculating the fidelities by assuming a theoretically expected value
% given by an independently measured $u$ as determined from the heralding
% efficiency would have very little effect on these fidelities.
We use an expected value for the bias parameter of $u=-0.96$, corresponding
to a heralding efficiency of 4\%, which is consistent with our measured
efficiencies. We note that the fidelities are not particularly sensitive to
changes in the expected value of $u$. If this value is 
varied between $-0.93$ and $-0.98$, the fidelities for ${\hat \Pi}_1$ in Table
\ref{tab:table4} are unchanged, while those of ${\hat \Pi}_2$ remain above
0.999.

\begin{figure}
    \centering
    \includegraphics{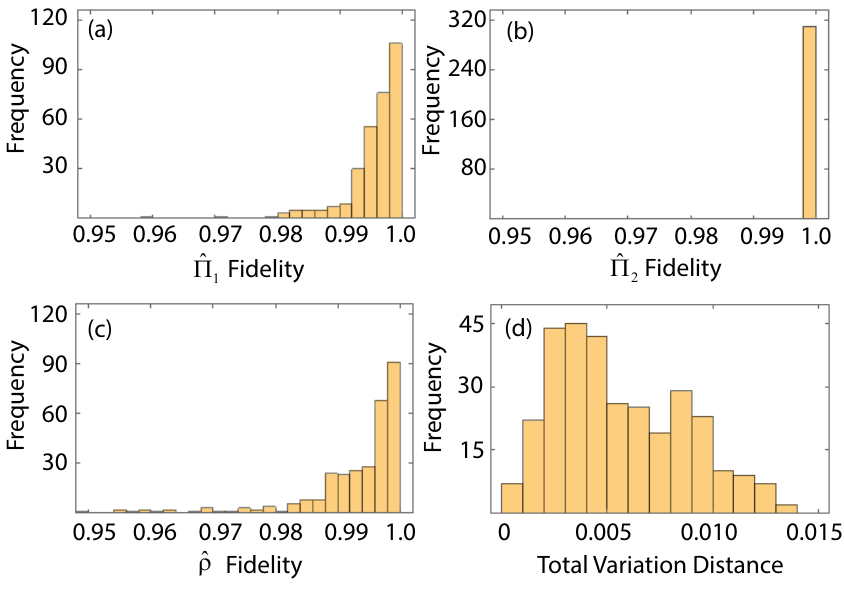}
    \caption{Histograms (a) and (b) show the
    fidelities of the theoretically expected ${\hat \Pi}_1$ and
    ${\hat \Pi}_2$ with the experimentally determined operators, (c)
    shows the fidelity of the theoretical and experimental $\hat \rho$,
    and (d) the total variation distance between the measured and the
    model probability distributions.
    This data was analyzed using the second detector model, as applied
    to $\rm{D_b}$.}
    \label{fig:hists2}
\end{figure}

Fig. \ref{fig:hists2} shows
histograms of the measured fidelities and TVDs corresponding
to all of the trials. One thing to note is
that despite the relatively low heralding efficiency, the reconstruction of
the density operator is nearly as good here as it was for the first
detector model.
The fidelity of $\hat \rho$ is over 0.99 for 76\% of the trials, and
over 0.98 for 92\% of trials. 

The fidelities for the POVMs are better for the second detector model
than they were for the first. For ${\hat \Pi}_1$ the fidelities exceed 0.99
for 89\% of the trials, and 0.98 for 97\% of trials. For
${\hat \Pi}_2$ the lowest fidelity is 0.999424. Clearly the large
imbalance is having a strong effect on the fidelities of ${\hat \Pi}_2$.
With $u \approx -1$, ${\hat \Pi}_2$ is approximately equal to the
identity operator and is largely independent of $\vec w$. 

Finally, the TVDs for the second detector model have an average value of 
$0.006 \pm 0.003$, so the model fits the data quite well. Note that this
measure is independent of any assumptions about the theoretically
expected state or POVMs. In particular, it is independent of 
the expected value of $u$.

\subsubsection{Discussion of the models}

Recall that the first detector model treats the apparatus in the box labeled 
``Measurement" in Fig.~\ref{fig:app} as a single, two-outcome POVM. 
The advantage of this model is that because the detections are post selected,
it is insensitive the heralding efficiency of the single-photon detections.
For this model
to be valid, it is necessary for $u$ and $\vec w$ to be the same for both 
$D_a$ and $D_b$. Important contributors to these parameters are the detector
efficiencies, and the ability of the PBS to accurately separate orthogonally
polarized photons.

If the detection apparatus does not satisfy these conditions,
it is necessary to use the
second detector model and determine separate POVMs for $D_a$ and
$D_b$. In
our experiments the low heralding efficiency meant that there was a significant
bias toward no detections, so the no-detection measurement operator
was nearly equal to the
identity operator, and hence largely independent of the measurement parameters.
The fidelity of the operator corresponding to a detection was found to be
somewhat 
insensitive to changes in the expected value of 
$u$, and hence to changes in the expected $w$. Thus, if the 
detection efficiency is low, one should be aware of these limitations 
in the reconstructed POVMs. This problem would be reduced or eliminated
in the case of higher detection efficiencies. Despite these issues, the
polarization state
is reconstructed with high fidelity, so one can still have high confidence in 
it, even with low efficiency detection.

\section{Conclusions}
We have described a technique which is capable of estimating both
the unknown quantum state of a single qubit, and a two-outcome POVM
that performs 
measurements of this qubit, in a self-consistent manner. This is done by
performing a series of known, unitary transformations between the state
preparation and measurement stages. This technique
makes minimal assumptions about
the state and the POVMs. We present two different models for the POVMs. In
one model we assume that the $u$ and $\vec w$ parameters of the two
detectors are the same,
but in the other we do not need this assumption.
% Due to inherent gauge degrees of freedom, the directions
% of the vectors $\vec p$ and $\vec w$ that describe the state and the POVM
% are determined to within a
% two-fold ambiguity. The magnitudes of these vectors is constrained, but
% there is also a continuous gauge degree of freedom that trades off between
% the purity of the state and the discriminating power of the detector. 

We assume that
the unitary transformations are known. In our experiments this assumption is
valid because the transformations are characterized
classically with high
fidelity, as demonstrated in Appendix \ref{append:pol}. While our assumption
will not be valid in all experiments, it will be valid in many optical
experiments where the transformations can be calibrated classically
\cite{carolan_2015,mennea_2018,flamini_2018}.
Knowing these
transformations is what allows us to self-consistently determine the state
and the POVMs with only ten measurements. This saves time when
compared to other
techniques that do not assume the transformations are known, but
require approximately 50 measurements \cite{blume-kohout_2013c,zhang_2020}.

We have experimentally implemented this technique, and applied
it to a system
described by the polarization of individual photons. We find that the
technique works quite well, as the fidelities between expected and measured
density operators and POVMs are found to exceed 0.98 for at least 92\% of
our 310 experimental trials. 

\begin{acknowledgments}
We wish to thank M. Schlosshauer and S. J. van Enk for helpful discussions.
This project was funded by NSF (1855174). AS acknowledges support, in part,
of a Reed College Science Research Fellowship. JMC acknowledges support,
in part,
by Galakatos Funds from Reed College. 
\end{acknowledgments}

\appendix

\section{\label{append:pol} Polarization transformations}
Here we describe our apparatus that generates rotations
$\tilde U\left[ {{{\vec k}\left ( \theta, \phi \right)},{\varphi}}  \right]$
in the Bloch sphere for polarization qubits.
% Note that for polarization 
% the Bloch sphere is isomorphic to the Poincar\'{e} sphere of classical
%optics.
% Since in this appendix we are treating these transformations classically,
% we will treat them as operating in the Poincar\'{e} sphere.
% the general polarization transformations
% $\tilde U\left[ {{{\vec k}\left ( \theta, \phi \right)},{\varphi}}  \right]$
% \cite{Bhandari_1998,Simon_1990,Sit_2017}. This is most easily seen from
% the fact that $\tilde U$ is specified by three independent parameters
% $(\theta, \phi, \varphi)$, while the half- and quarter-wave plate
% combination has only two independent parameters (the rotation angles
% of the two wave plates).
We describe the theory behind
the design of our device that implements these general polarization
transformations \cite{bhandari_1989,simon_1990,sit_2017},
and present experimental results that demonstrate
its performance.

\subsection{\label{sec:thry}Theory}

Polarization transformations that do not modify the total intensity are
unitary transformations, and may be represented by 3x3 matrices.
As seen in Fig.
\ref{fig:1}, a general polarization transformation is represented in the
Bloch sphere by a rotation ${\tilde R}\left( {\theta,\phi,\varphi} \right)$,
having a rotation axis $\vec k$ and a rotation angle $\varphi$. The
rotation axis is parameterized by two angles, $\theta$ and $\phi$. 

As seen in Fig. \ref{fig:1}, we take the rotation axis ${\vec k}$ to make
an angle of $\theta$ from the  2-axis ($\left| R \right\rangle $) in the
Bloch sphere, and its projection onto the plane perpendicular to this
axis to make an angle of $\phi$ from the  3-axis ($\left| H \right\rangle $).
With this convention, the rotation axis is given by Eq. (\ref{eq:a2}).

% \begin{equation}
% \vec k = \left( {\begin{array}{*{20}{c}}
% {{k_1}}\\
% {{k_2}}\\
% {{k_3}}
% \end{array}} \right) = \left( {\begin{array}{*{20}{c}}
% {\sin \left( {\theta} \right)\sin \left( {\phi} \right)}\\
% {\cos \left( {\theta} \right)}\\
% {\sin \left( {\theta} \right)\cos \left( {\phi} \right)}
% \end{array}} \right)
% 	\label{eq:pol:a}.
% \end{equation}
%
If we define $c = \cos \left( {\varphi} \right)$,
$d = 1 - \cos \left( {\varphi} \right)$,
$s = \sin \left( {\varphi} \right)$, we can express ${\tilde R}\left(
{\theta,\phi,\varphi} \right)$ in matrix form as

\begin{eqnarray}
& \tilde R \left( {\theta,\phi,\varphi} \right) = \nonumber\\
& \left( {\begin{array}{*{20}{c}}
{dk_1^2 + c}&{d{k_1}{k_2} - s{k_3}}&{d{k_3}{k_1} + s{k_2}}\\
{d{k_1}{k_2} + s{k_3}}&{dk_2^2 + c}&{d{k_3}{k_2} - s{k_1}}\\
{d{k_3}{k_1} - s{k_2}}&{d{k_3}{k_2} + s{k_1}}&{dk_3^2 + c}
\end{array}} \right)
	\label{eq:pol:b}.
\end{eqnarray}

Furthermore, let $R_{ij}$ be the element in the $i^{th}$ row and $j^{th}$
column of matrix $\tilde R$. Given a 3x3 unitary matrix that represents
a rotation, we can extract the rotation angle and
rotation axis using \cite{muga_2006}
\begin{subequations}
\label{eq:pol:d}
\begin{equation}
\cos \left( {\varphi} \right) = \frac{1}{2}\left[ {{\rm{Tr}}
\left( {\tilde R} \right) - 1} \right]
	\label{subeq:pol:a}
\end{equation}
\begin{equation}
{k_1} = \frac{1}{{2\sin \left( {\varphi} \right)}}\left( {{R_{32}}
- {R_{23}}} \right)
	\label{subeq:pol:b}
\end{equation}
\begin{equation}
{k_2} = \frac{1}{{2\sin \left( {\varphi} \right)}}\left( {{R_{13}}
- {R_{31}}} \right)
	\label{subeq:pol:c}
\end{equation}
\begin{equation}
{k_3} = \frac{1}{{2\sin \left( {\varphi} \right)}}\left( {{R_{21}}
- {R_{12}}} \right)
	\label{subeq:pol:d}.
\end{equation}
\end{subequations}

We wish to implement general polarization transformations, described by
Eq. (\ref{eq:pol:b}), using wave plates. Consider a wave plate that has
a phase shift $\phi_W$ between its fast and slow axes, and whose fast-axis
is rotated by $\theta_W$ from the horizontal. If we define $c ' = \cos
\left( {2\theta_W } \right)$ and $s ' = \sin \left( {2\theta_W } \right)$,
the transformation matrix that describes this wave plate can be written
as \cite{collett_2005}

\begin{eqnarray}
&\tilde M\left( {\theta_W ,\phi_W} \right) = \nonumber \\
&\left( {\begin{array}{*{20}{c}}
{{{s'}^2} + {{c'}^2}\cos \left( \phi_W  \right)}&{ -
c'\sin \left( \phi_W  \right)}&{c's'\left[ {1 - \cos
\left( \phi_W  \right)} \right]}\\
{c'\sin \left( \phi_W  \right)}&{\cos \left( \phi_W  \right)}
&{ - s'\sin \left( \phi_W  \right)}\\
{c's'\left[ {1 - \cos \left( \phi_W  \right)} \right]}&
{s'\sin \left( \phi_W  \right)}&{{{c'}^2} + {{s'}^2}\cos
\left( \phi_W  \right)}
\end{array}} \right) .\nonumber
%\label{eq:c}.
\end{eqnarray}
A special case that is of interest to us is the matrix that corresponds
to a quarter-wave plate, ${\tilde M_{Q}}\left( \theta_W  \right) =
\tilde M\left( {\theta_W ,\phi_W  = \pi /2} \right)$.

The wave plate implementation of a general polarization transformation that
we use is given by

\begin{eqnarray}
 &  \tilde R\left( {\theta,\phi,\varphi} \right) =\nonumber \\
 &  {\tilde M_{Q}}\left( {\frac{{\phi}}{2}} \right)\tilde M\left[
 {\frac{\pi }{4} + \frac{1}{2}\left( {\phi - \theta} \right),\varphi}
 \right]{\tilde M_{Q}}\left( {\frac{{\phi}}{2} + \frac{\pi }{2}} \right) 
   \label{eq:pol:e}.
\end{eqnarray}
Multiplying the matrices to the right of the equal sign in Eq.
(\ref{eq:pol:e}), and applying Eq. (\ref{eq:pol:d}) to the resulting
matrix, verifies that this combination of wave plates does indeed
implement $ \tilde R\left( {\theta,\phi,\varphi} \right)$. Physically, this
corresponds to a variable-wave plate placed between two quarter-wave
plates. From Eq. (\ref{eq:pol:e}) we find that the phase shift of the
variable-wave plate must be equal to the rotation angle in the Bloch
sphere, $\phi_W=\varphi$, and its rotation angle must be 
\begin{equation}
{\theta _W} = \frac{\pi }{4} + \frac{1}{2}\left( {\phi - \theta} \right) 
   \label{eq:pol:f}.
\end{equation}
The rotation angle of the second quarter-wave plate is ${\theta _Q} =
{\phi}/{2}$
, while the rotation angle of the first is $\theta_{Q}+ \pi/2$.

\subsection{Process Fidelity}
We verify the operation of our device by inputting classical light of known
polarization into it, and measuring three of the Stokes parameters
of the light that
emerges from it. (The fourth Stokes parameter is the total intensity, and
we normalize this to 1.)
The classical Stokes parameters are equivalent to the
parameters of the vector $\vec p$ that we use to describe the quantum 
state of polarization (although the numbering scheme for the two is
different: $p_1=S_2$, $p_2=S_3$ and $p_3=S_1$).

We wish to determine the ``classical fidelity" of our transformations.
Note that the theoretically expected Stokes vectors in our experiment
are ``pure": they have unit magnitude.
If one of
the states is pure, the quantum fidelity of Eq. (\ref{eq:p}) can be
simplified to
\cite{barnett_2009}
\begin{equation}
{F} = {\rm Tr}({\hat \rho}_1  {\hat \rho}_2)
   \label{eq:pol:f2}.
\end{equation}
Using Eq. (\ref{eq:a}), it is straightforward to demonstrate that
we can rewrite this expression in terms of the vectors that
describe the states as

\begin{equation}
{F} = \frac{1}{2}(1+{\vec p}_1 \cdot {\vec p}_2)
   \label{eq:pol:f3}.
\end{equation}
As such, we take the classical fidelity to be given by Eq.
(\ref{eq:pol:f3}), where ${\vec p}_1$ and ${\vec p}_2$ are the Stokes
vectors of the expected and measured polarizations.
%
% We can characterize the fidelity of the transformations that we perform by
% treating them as quantum processes. Despite the fact that in these
% experiments we use classical light to characterize the transformations, it
% is reasonable to consider them as quantum processes for this purpose. This
% is because the matrices that describe classical polarization transformations,
% and those that describe operations on the polarization states of individual
% photons, are the same. A single photon will be transformed in the same way
% that classical light will be.
%
% From the experimentally measured Stokes parameters we can calculate the
% equivalent density operator using Eq.~(\ref{eq:a}) (we are essentially
% performing tomography of the polarization state \cite{altepeter_2006}). We
% compare this to the theoretically expected density operator and determine
% the fidelity using Eq.~(\ref{eq:p}). 
%
The fidelity of the entire process that describes the 
transformation is given by the fidelity between the measured output state
and the theoretically expected output state, averaged over many states
\cite{poyatos_1997,magesan_2011}.

\subsection{Experiment}
\begin{figure}
\includegraphics[width=\columnwidth]{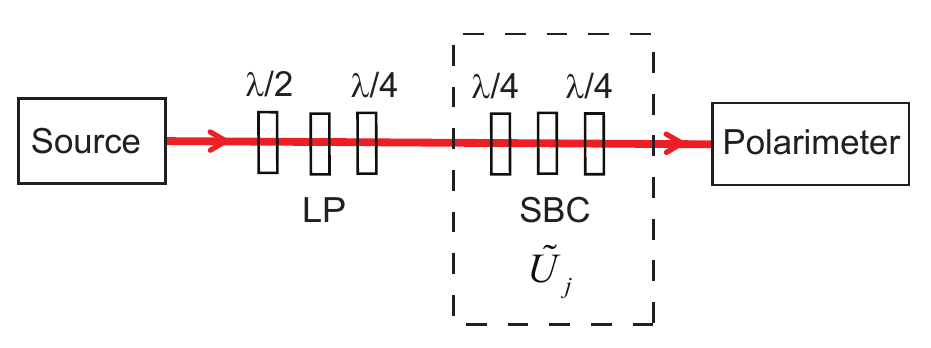}
% Here is how to import EPS art
\caption{\label{fig:2} The experimental apparatus for characterizing
our device that implements ${\tilde U}_j$. The source is a laser diode
coupled to a single-mode optical fiber. Here LP denotes a
linear polarizer, SBC denotes a Soleil-Babinet compensator,
$\lambda/2$ denotes a half-wave
plate and $\lambda/4$ denotes a quarter-wave plate. 
The two
quarter-wave plates on either side of the SBC are Berek compensators that 
have been adjusted for $\lambda/4$ retardation.}
\end{figure}

Our experimental apparatus is depicted in Fig. \ref{fig:2}. The light source
is an 808nm laser diode, coupled to a polarization-preserving, single-mode
fiber, which acts as a spatial filter.
We prepare the polarization that we input to our device by
rotating a linear polarizer and a quarter-wave plate. A half-wave plate
preceding the linear polarizer allows us to adjust the intensity. The beam
passes through our unitary-transformation apparatus, and the
polarization emerging
from it is analyzed by a commercial polarimeter (Thorlabs PAX1000IR1).

% As seen in Fig. \ref{fig:2}, measurements are performed by detecting the
% light emerging from the two outputs of a polarizing beam splitter (PBS).
% Quarter- and half-wave plates in front of the PBS determine what
% polarizations are measured by the detectors. Rotations of these wave
% plates allow us to measure the Stokes parameters of the light emerging
% from our device.  
%The rotation angles of the wave plates and the phase shift of the SBC that
% make up our device, and the rotation angles of the wave plates in the
% measurement stage, are all under computer control. 
%For example, we can enter the angles $(\theta_j, \phi_j, \varphi_j)$ that
% determine $\tilde U_j$, and the experimental apparatus is automatically
% adjusted to implement the proper transformation.

To illustrate how our device implements general polarization
transformations, we perform a series of measurements. We begin by fixing
the rotation angles of the wave plates in our device that implements
${\tilde U}_j$, which fixes the
rotation axis in the Bloch sphere. Next we fix the input polarization
by setting the rotation angles of the linear polarizer and quarter-wave
plate that follow our source. 
% Then we set the wave plates in our
% measurement device to project onto the  horizontal/vertical $(p_3)$ axis
% of the Bloch sphere. 
Now we scan the phase shift of the SBC (which
scans the rotation angle) in 17 equally-spaced steps that range from 0 to
$2\pi$ inclusive, while measuring the normalized Stokes vectors of the
output polarization.
% In this way we measure the $p_3$ projection of all 17
% rotations. Now we change the wave plates in the measurement apparatus to
% project onto the diagonal/anti-diagonal ($\pm 45^{\circ}$) axis, and
% repeat the scan of rotation angles. Finally, we set our measurement 
% apparatus to project onto the left-/right-circular axis, and again repeat
% the scan of rotation angles. 
These vectors should form a
“ring” around the rotation axis. We can now vary the input polarization, 
and repeat the sequence of measurements described above to sweep out
another ring. In order to obtain reproducible results, the rotation axes of
the three wave plates that make up our device are controlled by computer
via stepper motors.
The phase shift of the SBC is also computer controlled using a DC servo
motor.
\begin{figure}
\includegraphics[width=\columnwidth]{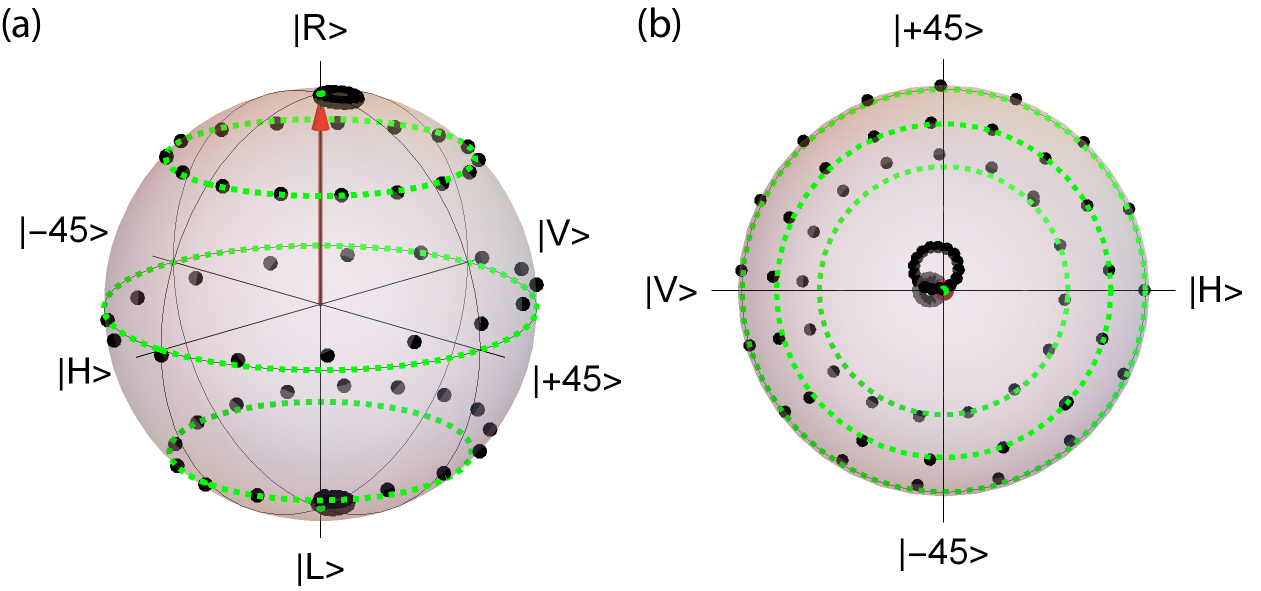}
% Here is how to import EPS art
\caption{\label{fig:3} Experimentally measured Stokes vectors (black dots)
and corresponding theoretical predictions (green lines). The rotation axis
is shown as a red arrow. (a) is a side view of the Bloch sphere, while
(b) shows a view looking down along the rotation axis.}
\end{figure}
\begin{figure}
\includegraphics[width=\columnwidth]{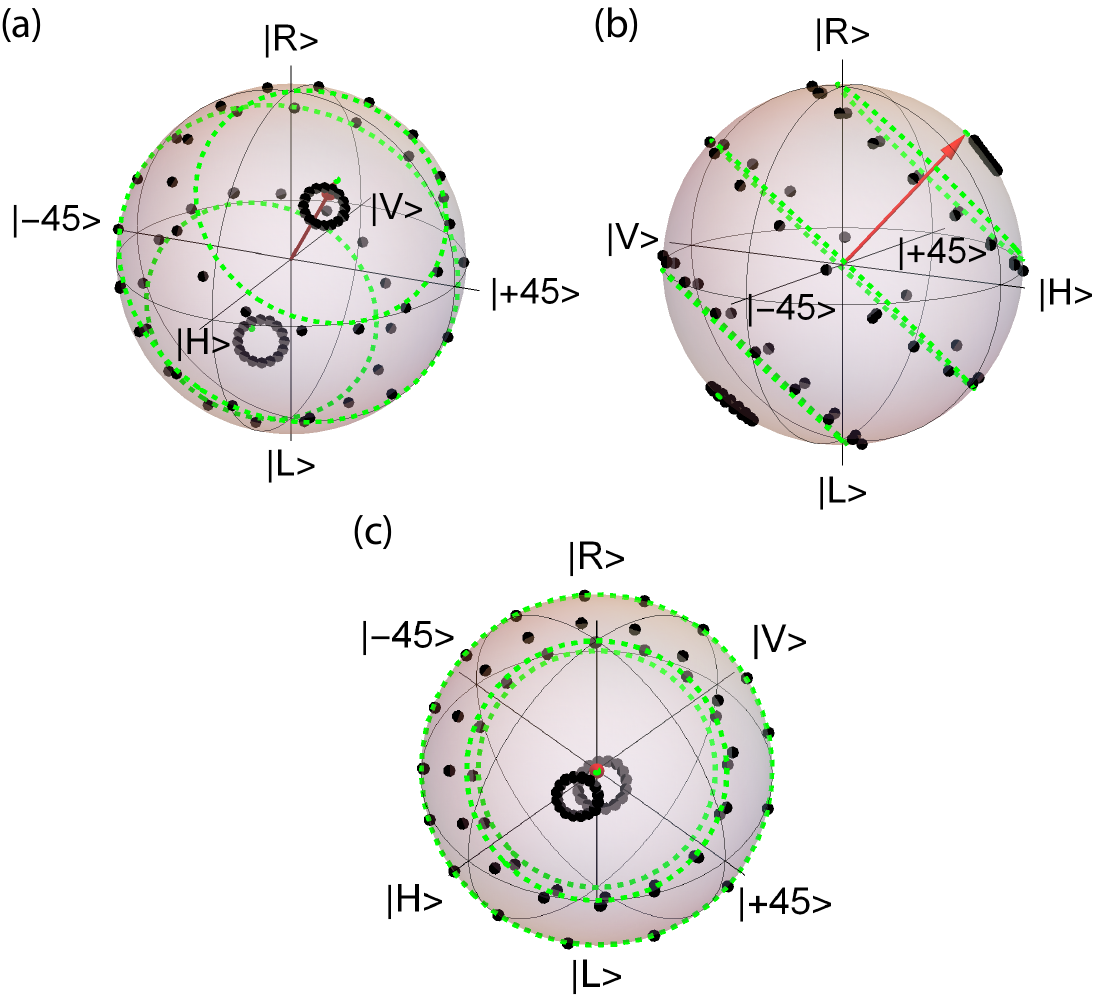}
% Here is how to import EPS art
\caption{\label{fig:4} Experimentally measured Stokes vectors (black dots)
and corresponding theoretical predictions (green lines). The rotation axis
is shown as a red arrow. (a) and (b) are side views of the Bloch sphere,
while (c) shows a view looking down along the rotation axis.}
\end{figure}

Figure \ref{fig:3} shows the result of measurements performed with our
device set to perform a rotation about the  $\left| R \right\rangle $ axis.
In this figure there are 5 different input polarizations, so there are 5
rings. The black dots show the experimentally measured Stokes vectors, while
the green lines represent the theoretical predictions. Figure \ref{fig:4}
shows the same thing, but for a rotation axis given by the angles
$\theta=\phi=\pi/4$. As expected, the experimental Stokes vectors all lie
near the surface of the Bloch sphere, which means that the measured
polarization states are nearly pure. The experimental points lie near the
theoretical curves, indicating at least qualitative agreement between the
theory and the experiments. 

To verify that the agreement is quantitative,
we can compute process fidelities from our measurements.
The data from the transformation displayed in Fig. \ref{fig:3} yield an
average process fidelity of $F=0.997\pm0.002$.
For the data depicted in Fig. \ref{fig:4} we find
$F=0.994\pm0.005$. The transformations
that we use in our experiments are listed in Table \ref{tab:5}. We
have determined the fidelities for each of
these and find that the mean process fidelity between the
theoretically expected and corresponding 
measured transformation is at least 0.994.
These fidelities are similar with those that we
obtained for the quantum density operators (e.g., Fig. \ref{fig:hists}). These
measurements are thus
consistent with the idea that the thing that currently limits the precision 
of
our measurements is our ability to control the unitary transformations.

In the tomographic reconstructions of Sec. 3.B 
we use the theoretically expected 
${\tilde U}_j$'s of Table \ref{tab:5}. We do this because
it is these transformations
that yield the linear inversion of Eq. (\ref{eq:g}), given below.
It might be possible
in the future to perform some
form of
``classical process tomography" from our classical calibration data to yield
maximum likely descriptions of our transformations, and this
might improve our results. But, as stated above,
the theoretical transformations we use in our reconstructions
have a process fidelity of at least
0.994 with the measured transformations, so a full process tomography
determination of the ${\tilde U}_j$'s cannot yield transformations
that are significantly more accurate than the ones we use.

To ensure that the classical calibration we perform here is still valid for
the quantum experiments, the apparatus labeled ${\tilde U}_j$ in
Fig.~\ref{fig:2} is left in place, and the beam path is
defined with irises.
The polarization preparation is then replaced by the single-photon
state preparation
apparatus in Fig.~\ref{fig:app}, which is 
mode-matched into the irises. The polarimeter is replaced by the 
measurement apparatus in Fig.~\ref{fig:app}

% For example the linear polarizer in Fig.~\ref{fig:2} is formed by using 
% the two BDPs in Fig.~\ref{fig:app}, and placing beam block in one of the
% two paths.

% mean that the device we have
% constructed implements transformations that are well described by the
% corresponding theoretically expected transformations.

% The 
% \textbf{[can we estimate the angle error, and does it agree with the
% 0.05 estimate above?]}

% We do not expect our theoretical predictions to be perfect. Errors
% in producing the initial polarization, or in performing the measurements
% of the Stokes vectors (due to imprecise rotation of the polarizer or wave
% plates, or imperfections in the wave plates) would mean that our predictions
% would not match the true experimental parameters. We thus conclude that the
% measured fidelities represent a lower bound on the actual performance of our
% device.
%
\section{\label{append:eq} Linear equations}

\begin{table}
\caption{\label{tab:5}The ten unitary transformations that we use in
our experiments.}
\begin{ruledtabular}
\begin{tabular}{cccc}
% &\multicolumn{2}{c}{$D_{4h}^1$}&\multicolumn{2}{c}{$D_{4h}^5$}\\
${\tilde U}_j$&$\theta_j$&$\phi_j$&$\varphi_j$\\ \hline
${\tilde U}_1$&0&0&0 \\
${\tilde U}_2$&0&0&$\pi/2$ \\
${\tilde U}_3$&0&0&$\pi$ \\
${\tilde U}_4$&$\pi/2$&0&$\pi$ \\
${\tilde U}_5$&$\pi/2$&0&$\pi/2$ \\
${\tilde U}_6$&$\pi/2$&$\pi/2$&$\pi/2$ \\
${\tilde U}_7$&$\pi/2$&$\pi/2$&$\pi$ \\
${\tilde U}_8$&$\pi/2$&$\pi/4$&$\pi$ \\
${\tilde U}_9$&$\pi/4$&0&$\pi$ \\
${\tilde U}_{10}$&$\pi/4$&$\pi/2$&$\pi$ \\

\end{tabular}
\end{ruledtabular}
\end{table}

Here we describe our solutions to the set of linear equations.

The ten Block-sphere rotations that we 
use for our measurements are listed in Table \ref{tab:5}.
Substituting these into Eq. (\ref{eq:d2})
yields
\begin{subequations}
\label{eq:f}
\begin{equation}
u + p_1w_1 + p_2w_2 + p_3w_3 = E_1
	\label{subeq:f:a}
\end{equation}
\begin{equation}
u + p_3 w_1 + p_2 w_2 - p_1 w_3 =  E_2
	\label{subeq:f:b}
\end{equation}
\begin{equation}
u - p_1 w_1 + p_2 w_2 - p_3 w_3 =  E_3
	\label{subeq:f:c}
\end{equation}
\begin{equation}
u - p_1 w_1 - p_2 w_2 + p_3 w_3  = E_4
	\label{subeq:f:d}
\end{equation}
\begin{equation}
u - p_2 w_1 + p_1 w_2 + p_3 w_3 = E_5
	\label{subeq:f:e}
\end{equation}
\begin{equation}
u + p_1 w_1 - p_3 w_2 + p_2 w_3  = E_6
	\label{subeq:f:f}
\end{equation}
\begin{equation}
u + p_1 w_1 - p_2 w_2 - p_3 w_3  = E_7
	\label{subeq:f:g}
\end{equation}
\begin{equation}
u + p_3 w_1 - p_2 w_2 + p_1 w_3  = E_8
	\label{subeq:f:h}
\end{equation}
\begin{equation}
u - p_1 w_1 + p_3 w_2 + p_2 w_3  = E_9
	\label{subeq:f:i}
\end{equation}
\begin{equation}
u + p_2 w_1 + p_1 w_2 - p_3 w_3  = E_{10}
	\label{subeq:f:j}.
\end{equation}
\end{subequations}
%\end{singlespace}
%
%The expectation values $B_j$'s are quantities that can be measured in
%experiments, while $u$ and the $w_i$'s and $p_i$'s are
% what we are trying to determine.
% Now define $x_{ij}=p_iw_j$ for $i,j=1,2,3$. These 9 parameters represent all
% of the products of the components of $\vec p$ and $\vec w$.
%
% \begin{subequations}
% \label{eq:g2}
% \begin{equation}
% x=w_1p_1
% \end{equation}
% \begin{equation}
% y=w_2p_2 
% \end{equation}
% \begin{equation}
% z=w_3p_3. 
% \end{equation}
% \end{subequations}
%
Substituting $x_{ij}=p_i w_j$ into Eq. (\ref{eq:f}), we see that
Eq. (\ref{eq:f})
represents a set of 10 linear equations in 10 unknowns
$u, x_{ij}\: (i,j=1,2,3)$. Solving
them yields

\begin{subequations}
\label{eq:g}
\begin{equation}
u = \frac{1}{4}( E_1 + E_3 + E_4 + E_7)
	\label{subeq:g:a}
\end{equation}
\begin{equation}
x_{11} = \frac{1}{4}(E_1 - E_3 - E_4 + E_7)
	\label{subeq:g:b}
\end{equation}
\begin{equation}
x_{12} = \frac{1}{4} (-E_1  - E_3 - E_4 + 2 E_5 - E_7 + 2 E_{10})
	\label{subeq:g:c}
\end{equation}
\begin{equation}
x_{13} = \frac{1}{4}(E_1 - 2 E_2 + E_3 - E_4 - E_7 + 2 E_8)
	\label{subeq:g:d}.
\end{equation}
\begin{equation}
x_{21} = \frac{1}{4}(E_1  - E_3 + E_4 - 2 E_5 - E_7 + 2 E_{10})
	\label{subeq:g:e}.
\end{equation}
\begin{equation}
x_{22} = \frac{1}{4}(E_1 + E_3 - E_4 - E_7)
	\label{subeq:g:f}.
\end{equation}
\begin{equation}
x_{23} = \frac{1}{4}(-E_1 - E_3 - E_4 + 2 E_6 - E_7 + 2 E_9)
	\label{subeq:g:g}.
\end{equation}
\begin{equation}
x_{31} = \frac{1}{4}(-E_1 + 2 E_2 - E_3 - E_4 - E_7 + 2 E_8)
	\label{subeq:g:h}.
\end{equation}
\begin{equation}
x_{32} = \frac{1}{4}(E_1 - E_3 - E_4 - 2 E_6 + E_7 + 2 E_9)
	\label{subeq:g:i}.
\end{equation}
\begin{equation}
x_{33} = \frac{1}{4}(E_1 - E_3 + E_4 - E_7)
	\label{subeq:g:j}.
\end{equation}
\end{subequations}
Thus, performing measurements of expectation values 
with the settings given in Table
\ref{tab:5} and using Eq. (\ref{eq:g}) yields values for the 
quantities $u$, $x_{ij}$.

Note that the settings given in Table
\ref{tab:5} are not unique. Any ten settings that yield linearly independent
equations for the expectation values [Eq. (\ref{eq:f})] will allow us
to solve for 
$u$ and the $x_{ij}$'s. We have chosen these particular settings because
the solutions in this case [Eq. (\ref{eq:g})] are fairly simple.

\end{document}